\documentclass[letterpaper,11pt,onecolumn, nodraft]{IEEEtran}
\usepackage{graphicx,latexsym,epsfig,amssymb,amsmath,subfigure}
\usepackage{setspace}
\doublespace
\begin{document}

\title{Pricing and Investments in Internet Security \\ \emph{A Cyber-Insurance Perspective}}


\author{Ranjan Pal,~\IEEEmembership{Student Member,~IEEE,}
        Leana Golubchik,~\IEEEmembership{Member,~IEEE,}
\thanks{R. Pal and L. Golubchik are with the Department
of Computer Science, University of Southern California,
CA, 90089 USA. e-mail: \{rpal, leana\}@usc.edu.}}
\maketitle
\begin{abstract}
Internet users such as individuals and organizations are subject to different types of epidemic risks such as worms, viruses, spams, and botnets. To reduce the probability of risk, an Internet user generally invests in traditional security mechanisms like anti-virus and anti-spam software, sometimes also known as \emph{self-defense} mechanisms. However, such software does not completely eliminate risk. Recent works have considered the problem of residual risk elimination by proposing the idea of \emph{cyber-insurance}. In this regard, an important research problem is the analysis of optimal user self-defense investments and cyber-insurance contracts under the Internet environment.

In this paper, we investigate \emph{two} problems and their relationship: 1) analyzing optimal self-defense investments in the Internet, under \emph{optimal} cyber-insurance coverage, where optimality is an insurer objective and 2) designing optimal cyber-insurance contracts for Internet users, where a contract is a (premium, coverage) pair. By the term `self-defense investment', we mean the monetary-cum-precautionary cost that each user needs to invest in employing risk mitigating self-defense mechanisms, \emph{given} that it is optimally insured by Internet insurance agencies. We propose 1) a general mathematical framework by which co-operative and non-co-operative Internet users can decide whether or not to invest in self-defense for ensuring both, individual and social welfare and 2) models to evaluate optimal cyber-insurance contracts in a single cyber-insurer setting. Our results show that co-operation amongst users results in more efficient self-defense investments than those in a non-cooperative setting, under full insurance coverage, in an ideal single insurer cyber-insurance market, whereas in non-ideal single insurer markets of non-cooperative users, partial insurance driven self-defense investments are optimal. We also show the \emph{existence} of a cyber-insurance market in a single cyber-insurer scenario.

\emph{Keywords:} cyber-insurance, self-defense investments, information asymmetry
\end{abstract}
%
\IEEEpeerreviewmaketitle
\section{Introduction} \label{sec-intro}
The Internet has become a fundamental and an integral part of our daily lives. Billions of people nowadays are using the Internet for various types of applications. However, all these applications are running on a network, that was built under assumptions, some of which are no longer valid for today's applications, e,g., that all users on the Internet can be trusted and that there are no malicious elements propagating in the Internet. On the contrary, the infrastructure, the users, and the services offered on the Internet today are all subject to a wide variety of risks. These risks include denial of service attacks, intrusions of various kinds, hacking, phishing, worms, viruses, spams, etc. In order to counter the threats posed by the risks, Internet users\footnote{The term `users' may refer to both, individuals and organizations.} have traditionally resorted to antivirus and anti-spam softwares, firewalls, and other add-ons to reduce the likelihood of being affected by threats. In practice, a large industry (companies like \emph{Norton, Symantec, McAfee,} etc.) as well as considerable research efforts are centered around developing and deploying tools and techniques to detect threats and anomalies in order to protect the Internet infrastructure and its users from the negative impact of the anomalies.

In the past one and half decade, protection techniques from a variety of computer science fields such as cryptography, hardware engineering, and software engineering have continually made improvements. Inspite of such improvements, recent articles by Schneier \cite{sch} and Anderson \cite{ranr}\cite{amr} have stated that it is impossible to achieve a 100\% Internet security protection. The authors attribute this impossibility primarily to four reasons: 1) new viruses, worms, spams, and botnets evolve periodically at a rapid pace and as result it is extremely difficult and expensive to design a security solution that is a panacea for all risks, 2) the Internet is a distributed system, where the system users have divergent security interests and incentives, leading to the problem of `misaligned incentives' amongst users. For example, a rational Internet user might well spend \$20 to stop a virus trashing its hard disk, but would hardly have any incentive to invest sufficient amounts in security solutions to prevent a service-denial attack on a wealthy corporation like an Amazon or a Microsoft \cite{varian}. Thus, the problem of misaligned incentives can be resolved only if liabilities are assigned to parties (users) that can best manage risk, 3) the risks faced by Internet users are often correlated and interdependent. A user taking protective action in an Internet like distributed system creates positive externalities \cite{hh} for other networked users that in turn may discourage them from making appropriate security investments, leading to the `free-riding' problem \cite{gccr}\cite{jaw}\cite{mybm}\cite{oom}, and 4) network externalities affect the adoption of technology. Katz and Shapiro \cite{kschr} have analyzed that externalities lead to the classic S-shaped adoption curve, according to which slow early adoption gives way to rapid deployment once the number of users reaches a critical mass. The initial deployment is subject to user benefits exceeding adoption costs, which occurs only if a minimum number of users adopt a technology; so everyone might wait for others to go first, and the technology never gets deployed. For example, DNSSEC, and S-BGP are secure protocols that have been developed to better DNS and BGP in terms of security performance. However, the challenge is getting them deployed by providing sufficient internal benefits to adopting firms.

In view of the above mentioned inevitable barriers to 100\% risk mitigation, the need arises for alternative methods of risk management in the Internet. Anderson and Moore \cite{amr} state that microeconomics, game theory, and psychology will play as vital a role in effective risk management in the modern and future Internet, as did the mathematics of cryptography a quarter century ago. In this regard, \emph{cyber-insurance} is a psycho-economic-driven risk-management technique, where risks are transferred to a third party, i.e., an insurance company, in return for a fee, i.e., the \emph{insurance premium}. The concept of cyber-insurance is growing in importance amongst security engineers. The reason for this is three fold: 1) ideally, cyber-insurance increases Internet safety because the insured increases self-defense as a rational response to the reduction in insurance premium \cite{kmy1}\cite{kmy2}\cite{bs}\cite{yd}. This fact has also been mathematically proven by the authors in \cite{leb3}\cite{leb}, 2) in the IT industry, the mindset of `absolute protection' is slowly changing with the realization that absolute security is impossible and too expensive to even approach while adequate security is good enough to enable normal functions - the rest of the risk that cannot be mitigated can be transferred to a third party \cite{kmy3}, and 3) cyber-insurance will lead to a market solution that will be aligned with economic incentives of cyber-insurers and users (individuals/organizations) - the cyber-insurers will earn profit from appropriately pricing premiums, whereas users will seek to hedge potential losses. In practice, users generally employ a simultaneous combination of retaining, mitigating, and insuring risks \cite{bs2}.

Sufficient evidence exists in daily life (e.g., in the form of
auto and health insurance) as well as in the academic literature
(specifically focused on cyber-insurance)
\cite{kmy1}\cite{kmy2}\cite{leb3}\cite{leb}\cite{bs}
that insurance-based solutions are useful approaches to pursue,
i.e., as
a complement to other security measures (e.g., anti-virus software).
However, cyber-insurance has not yet become
a reality due to a number of unresolved research
challenges as well as practical considerations (as detailed below).
A number of these challenges are rooted in the differences between
cyber-insurance and other forms of insurance.
Specifically, these include:
\begin{itemize}
   \item {\em Networked environment.}
	The operation of systems and applications in a networked
	environments leads to new insurance challenges.  Specifically,
	the network's topology, node connectivity, form of interaction
	among the nodes, all lead to subsequent risk propagation
	characteristics.
	This in turn implies that considerations of
	interdependent security and correlated risk (among system
	participants) are significantly more complex in an Internet-type
	environment.  All this leads to challenges in modeling
	of network topologies, risk arrival, attacker models, and so on.
   \item {\em Information asymmetry.}
	Information asymmetry has a significant effect on most insurance
	environments, where typical considerations include inability to
	distinguish between users of different types as well as users
	undertaking actions that affect loss probability after the insurance
	contract is signed.  However, there are important aspects of
	information asymmetry that are particular to cyber-insurance.
	These include users hiding information from insurers, users
	lacking information aout networked nodes, as well as
	insurers lacking information about and not differentiating
	based on products (e.g., anti-virus software) installed by users.
	All this leads to challenges in modeling insurers and insured
	entities.
\end{itemize}
In this paper, we address the problem of pricing and investments in Internet security related to cyber-insurance-driven risk management under a correlated, interdependent, and information asymmetric Internet environment. Our problem is important because 1) for cyber-insurance to be popular amongst Internet users, a market for it should first exist, which in turn depends on the prices charged by the cyber-insurer (supply side) to its clients (demand side) and the subsequent profits earned and 2) once a market for cyber-insurance exists, Internet users would want to invest optimally in self-defense investments, given insurance coverage, so as to improve overall security. Optimal user investments is important for two reasons: 1) investing in self-defense mechanisms reduces a user's probability of facing risk. Given that a user has cyber-insurance coverage, increase in user self-defense investments reduces its premium charged by the cyber-insurer. Thus, its important to characterize the \emph{appropriate} amounts of investments by a user in self-defense, as well as in cyber-insurance, such that it maximizes its utility and 2) many distributed Internet applications like peer-to-peer file sharing, multicasting, and network resource sharing encourage co-operation between users to improve overall system performance. In  regard to security investments, cooperation invites an opportunity for a user to benefit from the positive externality\footnote{ An externality is a positive (external benefit) or negative(external cost) impact on a user not directly involved in an economic transaction.} that its investment poses on the other users in the network. However, its not evident that users invest better when they cooperate compared to when they do not, in regard to the network achieving greater overall security. In this paper, we want to study whether security investments are more efficient under cooperation than under non cooperation when it comes to achieving better overall network security.

We make the following research contributions in this paper. Before stating them, we emphasize that they are based on the expected utility theory model by von-Neumann and Morgenstern, which is the most widely used theory for analyzing micro-economic models. We also assume in all our models the presence of only one cyber-insurer providing service to its clients (Internet users).
\begin{enumerate}
\item We quantitatively analyze an $n$-agent model, using \emph{botnet} risks as a representative application,
and propose a general mathematical framework through which Internet users can decide 1) whether to invest and 2) how much to invest in self-defense mechanisms, \emph{given }that each user is optimally insured w.r.t. insurer objectives in perfect single insurer cyber-insurance markets(see Section \ref{sec-mfr}). Our framework entails each Internet user to invest optimally in self-defense mechanisms in order to improve overall network security, and is applicable to all risk types that inflict direct and/or indirect losses to users.
\item For ideal\footnote{An insurance environment with no information asymmetry between the cyber-insurer and the insured.} single insurer cyber-insurance markets, we perform a mathematical comparative study to show that cooperation amongst Internet users results in better self-defense investments w.r.t. improving overall network security when the risks faced by the users in the Internet are interdependent (see Section \ref{sec-cs}). We use basic concepts from both, cooperative and non cooperative game theory to support the claims we make in Sections \ref{sec-mfr} and \ref{sec-cs}. Our results are applicable to both, co-operative (e.g., distributed file sharing) as well as non-cooperative Internet applications, where in both application types a user has the option to be either co-operative or non-cooperative with respect to security parameters.
\item We derive optimal cyber-insurance contracts (\emph{(premium, coverage)} pairs) between the cyber-insurer and the insured under both, ideal as well as non-ideal cyber-insurance environments, and show that a market for cyber-insurance exists when there is a single cyber-insurer providing insurance to all Internet users (see Section \ref{sec-optcnr}. While existing literature show that information asymmetries leads to market failure, using \emph{mechanism design} theory, we design robust cyber-insurance contracts that account for information asymmetries, maximize cyber-insurer profits, and are in market equilibrium.

\end{enumerate}
Through our contributions, we jointly address an economics problem of both, the supply side (cyber-insurer) as well as the demand side (cyber-insured) and study the relationship between the two, i.e., we study the effect that prices set in a cyber-insurance contract has on the self-defense investment of an Internet user. For ease of presentation, we first address the investment problem of Internet user under a given cyber-insurance contract followed by the problem of pricing optimal cyber-insurance contracts. We do this because cyber-insurers are the first movers and account for optimal self-defense investments of Internet users when designing optimal insurance contracts.

\section{Related Work}
The field of cyber-insurance in networked environments has been triggered by recent results on the amount of individual user self-defense investments in the presence of network externalities. The authors in \cite{gccr}\cite{jaw}\cite{leb5}\cite{leb4}\cite{mybm}\cite{oom} mathematically show that Internet users invest too little in self-defense mechanisms relative to the socially efficient level, due to the presence of network externalities. These works just highlight the role of positive externalities in preventing users for investing optimally in self-defense investments. Thus, the challenge to improving overall network security lies in incentivizing end-users to invest in sufficient amount of self-defense investments inspite of the positive externalities they experience from other users in the network. In response to the challenge, the works in \cite{leb5}\cite{leb4} modeled network externalities and showed that a tipping phenomenon is possible, i.e., in a situation of low level of self-defense, if a certain fraction of population decides to invest in self-defense mechanisms, it could trigger a large cascade of adoption in security features, thereby strengthening the overall Internet security. However, they did not state how the tipping phenomenon could be realized in practice. In a series of recent works \cite{leb3}\cite{leb}, Lelarge and Bolot have stated that under conditions of no \emph{information asymmetry} \cite{wik}\cite{hv} between the insurer and the insured, cyber-insurance \emph{incentivizes} Internet user investments in self-defense mechanisms, thereby paving the path to trigger a cascade of adoption. They also show that investments in both self-defense mechanisms and insurance schemes are quite inter-related in maintaining a socially efficient level of security on the Internet.

Inspite of Lelarge and Bolot proposing the role of cyber-insurance for networked environments in incentivizing increasing user security investments, its common knowledge that the market for cyber-insurance has not blossomed with respect to its promised potential. Most recent works \cite{ssfw}\cite{rabohme} have attributed the underdeveloped market for cyber-insurance due to 1. \emph{interdependent security}, 2. \emph{correlated risk}, and 3. \emph{information asymmetries}. Thus, the need of the hour is to develop cyber-insurance solutions simultaneously targeting these three issues and identify other factors that might play an important role in promoting a developed cyber-insurance market. The works in \cite{rkk}\cite{leb3}\cite{leb} \cite{hoffman} touch upon the notion of information asymmetry and the effect it has on the insurance parameters, however none of the works explicitly model information asymmetry. In relation to tackling information asymmetry, the authors in \cite{ssfw}\cite{hoffman}\cite{leb3} propose the concept of premium differentiation and fines, but none of the works provide an analytical model to strengthen their point.  In addition, no work considers the cooperative and non cooperative nature of network users and the effect this has on the overall level of security and appropriate self-defense investments.

\section{A Mathematical Framework for Self-Defense Investments} \label{sec-mfr}
In this section, we propose a general mathematical framework for deciding on the appropriate self-defense investment of an Internet user, under \emph{optimal} cyber-insurance coverage, in ideal single insurer cyber-insurance markets. Here, we assume that Internet users could buy insurance from entities like Internet service providers (ISPs) to cover the risks posed by botnets\footnote{Cyber-insurance providers could also be third-party agencies other than ISPs or the government.}. For instance, the coverage could be in the form of money or protection against lost data/reputation. Our framework is applicable to direct/indirect risks, those that are caused by worms, viruses, and botnets. Direct risks result when threats such as worms, viruses, and botnets infect machines (computing device) that lack a security feature, whereas indirect losses result due to the contagion process of one machine getting infected by its neighbors.
\subsection{Model Description} \label{sec-md}
We consider $n$ identical\footnote{We assume identical users to ensure tractable analyses.} rational risk-averse users in a network, i.e., $E(U(w)) < U(E(w))$, where $w$ is the wealth possessed by a user. We assume the users to be cooperative to a variable degree, i.e, the network supports Internet applications where users cooperate with other users in some capacity with the intention to improve overall system performance but may or may not cooperate entirely. The users could either voluntarily cooperate by sharing information with other network users regarding self-defense investments, or be bound to cooperate due to a network regulation, which requires participating users to share self-defense investment information. The users may also decide not to cooperate at all depending on the nature of applications. Each user has initial wealth $w_{0}$ and is exposed to a substantial risk of size $R$ with a certain probability $p_{0}$.
(Here, risk represents the negative wealth accumulated by a user when it is affected by Internet threats.)

A user investing in self-defense mechanisms reduces its risk probability. For an amount $x$, invested in self-defense, a user faces a risk probability of $p(x)$, which is a continuous and twice differentiable decreasing function of investment, i.e., $p'(x) < 0$, $p''(x) > 0$, $lim_{x\rightarrow \infty}p(x) = 0$, and $lim_{x\rightarrow \infty}p'(x) = 0$. The investment $x$ is a function of the amount of security software the user buys and the effort it spends on maintaining security settings on its computing device.
   In addition to investing in self-defense mechanisms, a user either finds it optimal to buy either \emph{full} or \emph{partial} cyber-insurance coverage at a particular premium to eliminate its residual risk. The premium and coverage applicable to users are determined through optimal cyber-insurance contracts that we will investigate in Section \ref{sec-optcnr}. A user \emph{does not} buy insurance for high probability low risk events because 1) these events are extremely common and does not cause sufficient damage to demand insurance solutions and 2) the insurance company also has reservations in insuring every kind of risk for profit purposes. We also assume for the moment that there exists markets for cyber-insurance, i.e., cyber-insurance strengthens overall network security and there exists cyber-insurance contracts that are in market equilibrium. We will show in Section \ref{sec-optcnr} that markets can be made to exist for single-insurer cyber-insurance environments.

An Internet user apart from being directly affected by threats may be indirectly infected by the other Internet users. We denote the indirect risk facing probability of a user $i$ as $q(\overrightarrow{x}_{-i}, n)$, where $\overrightarrow{x}_{-i} = (x_{1},......,x_{i-1}, x_{i+1},....,x_{n})$ is the vector of self-defense investments of users other than $i$. An indirect infection spread is either `perfect' or `imperfect' in nature. In a perfect spread, infection spreads from a user to other users in the network with probability 1, whereas in case of imperfect spread, infection spreads from a user to others with probability less than 1. For a perfect information spread $q(\overrightarrow{x}_{-i}, n) = 1 - \prod_{j = 1, j\neq i}^{n} (1 - p(x_{j}))$, whereas in the case of imperfect spread, $q(\overrightarrow{x}_{-i}, n) < 1 - \prod_{j = 1, j\neq i}^{n} (1 - p(x_{j}))$. In this paper, we consider perfect spread only,
without loss of generality because the probability of getting infected by others in the case of imperfect spread is less than that in the case of perfect spread, and as a result this case is subsumed by the results of the perfect spread case.
%
%
Under perfect spread, the risk probability of a user $i$ is given as
\begin{equation}
        p(x_{i}) + (1 - p(x_{i}))q(\overrightarrow{x}_{-i}, n) = 1 - \prod_{j = 1}^{n}(1 - p(x_{j}))
\end{equation}
   and its expected final wealth upon facing risk is denoted as $w_{0} - x_{i} - (1 - \prod_{j = 1}^{n}(1 - p(x_{j}))\cdot IC) - R + IC$, where $(1 - \prod_{j = 1}^{n}(1 - p(x_{j}))\cdot IC$ is the premium and $IC$ denotes the insurance coverage\footnote{For full insurance coverage $R = IC$.}. The aim of a network user is to invest in self-defense mechanisms in such a manner so as to either maximize its expected utility of final wealth, or maximize the expected utility of net wealth in the network system, depending on the nature of the application.
\subsection{Mathematical Framework for Full Insurance Coverage} \label{sec-mfic}
In this section, we assume full cyber-insurance coverage and propose a general mathematical framework for deciding on the appropriate self-defense investment of an Internet user. It has been proved in \cite{wetzstein} that under fair premiums and in ideal insurance environments, a user finds its optimal to buy full coverage. In other situations, a user might buy full coverage but it might not be optimal for itself as it may end up paying unfair premiums to the insurer, who does not want to make negative profits. Thus, we assume here that full coverage is optimal for users under ideal cyber-insurance environments, given that users would only want to be charged fair premiums.

We model the following risk management scenarios: (1) users do not cooperate and do not get infected by other users in the network, (2) users cooperate and may get infected by other users in the network, (3) users do not cooperate but may get infected by other users in the network, and (4) users cooperate but do not get infected by other users in the network. We note that Case 4 is a special case of Case 2 and thus is subsumed in the results of Section \ref{sec-cb}. Scenarios 2 and 3 are realistic in the Internet where risks do spread even though applications may or may not allow co-operation. Scenarios 1 and 4 are idealistic cases and are analyzed for pathological reasons as well as for purposes of comparison with scenarios 2 and 3 w.r.t. optimal self-defense investments.
\subsubsection{Case 1: No Cooperation, No Infection Spread} \label{sec-ca}
Under full insurance, the risk is equal to the insurance coverage, and users determine their optimal amount of self-defense investment by maximizing their level of final wealth, which in turn is equivalent to maximizing their expected utility of wealth \cite{eb}. We can determine the optimal amount of self-defense investment for each user $i$ by solving for the value of $p$ that maximizes the following constrained optimization problem:
   \[arg max_{x_{i}} FW_{i}(x_{i}) = w_{0} - x_{i} - p(x_{i})R - R + IC\]
   or \[arg max_{x_{i}} FW_{i}(x_{i}) = w_{0} - x_{i} - p(x_{i})R\]
   subject to \[0 \leq p(x_{i}) \leq p_{0},\]
   where $FW_{i}$ is the final wealth of user $i$ and $p(x_{i})R$ is the premium for full insurance coverage. Taking the first and second derivatives of $FW_{i}$ with respect to $x_{i}$, we obtain
   \begin{equation}
   FW'_{i}(x_{i}) = -1 - p'(x_{i})R
   \end{equation}
   and
   \begin{equation}
   FW''_{i}(x_{i}) = - p''(x_{i})R < 0
   \end{equation}
   Thus, our objective function is globally concave. Let $x_{i}^{opt}$ be the optimal $x_{i}$ obtained by equating the first derivative to $0$. Thus, we have:
   \begin{eqnarray}
   p'(x_{i}^{opt})R = -1.
   \label{eqn:invest}
   \end{eqnarray}

   \emph{Economic Interpretation:} The left hand side (LHS) of Equation~(\ref{eqn:invest})
is the marginal benefit of investing an additional dollar in self-protection mechanisms, whereas the right hand side (RHS) denotes the marginal cost of the   investment. A user equates the LHS with the RHS to determine its self-defense investment.

   \emph{Conditions for Investment:} We first investigate the boundary costs. The user will not consider investing in self-defense if $p'(0)R \geq -1$ because its marginal cost of investing in any defense mechanism, i.e., -1, will be relatively equal to or lower than the marginal benefit when no investment occurs. In this case, $x_{i}^{opt} = 0$. If the user invests such that it has no exposure to risk, $x_{i}^{opt} = \infty$. When $p'(0)R < -1$, the costs do not lie on the boundary, i.e., $0 < x_{i}^{opt} < \infty$, and the user invests to partially eliminate risk (see Equation~(\ref{eqn:invest})).
\subsubsection{Case 2: Cooperation, Infection Spread} \label{sec-cb}
Under full insurance coverage, user $i$'s expected final wealth is given by
\begin{equation}
   FW_{i} = FW(x_{i},\overrightarrow{x}_{-i}) = w_{0} - x_{i} - (1 - \prod_{j = 1}^{n}(1 - p(x_{j})))R
\end{equation}
When Internet users co-operate, they jointly determine their optimal self-defense investments. We assume that co-operation and bargaining costs are nil. In such a case, according to Coase theorem \cite{ctm}, the optimal investments for users are determined by solving for the socially optimal investment values that maximize the aggregate final wealth (AFW) of all users. Thus, we have the following constrained optimization problem:
   \[arg max_{x_{i}, \overrightarrow{x}_{-i}} AFW = nw_{0} - \sum_{i = 1}^{n} x_{i} - n(1 - \prod_{j = 1}^{n}(1 - p(x_{j})))R\]
   \[0 \leq p_{i}(x_{i}) \leq p_{0},\,\forall i\]
   Taking the first and the second partial derivatives of the aggregate final wealth with respect to $x_{i}$, we obtain
   \begin{equation}
   \frac{\partial}{\partial x_{i}}(AFW) = - 1 - np'(x_{i})\prod_{j = 1, j \ne i}^{n}(1 - p(x_{j}))R
   \end{equation}
   and
   \begin{equation}
   \frac{\partial^{2}}{\partial x_{i}^2}(AFW) = -np''(x_{i})\prod_{j = 1,j \ne i}^{n}(1 - p(x_{j}))R < 0
   \end{equation}
   The objective function is globally concave, which implies the existence of a unique solution $x^{opt}_{i}(\overrightarrow{x}_{-i})$, for each $\overrightarrow{x}_{-i}$. Our maximization problem is symmetric for all $i$, and thus the optimal solution is given by $x^{opt}_{i}(\overrightarrow{x^{opt}_{-i}}) = x^{opt}_{j}(\overrightarrow{x^{opt}_{-j}})$ for all $j = 2,....,n$. We obtain the optimal solution by equating the first derivative to zero, which gives us the following equation
   \begin{eqnarray}
   np'(x_{i}^{opt}(\overrightarrow{x}_{-i}))\prod_{j = 1, j \ne i}(1 - p(x_{i}))R = - 1
   \label{eqn:invest2}
   \end{eqnarray}

\emph{Economic Interpretation:} The left hand side (LHS) of Equation~(\ref{eqn:invest2}) is the marginal benefit of investing in self-defense. The right hand side (RHS) of Equation~(\ref{eqn:invest2}) is the marginal cost of investing in self-defense, i.e., -1. We obtain the former term of the marginal benefit by internalizing the positive externality\footnote{Internalizing a positive externality refers to rewarding a user, who contributes positively and without compensation, to the well-being of other users, through its actions.}, i.e., by accounting for the self-defense investments of other users in the network. The external well-being posed to other users by user $i$ when it invests an additional dollar in self-defense is $-p'(x_{i})\prod_{j = 1, j \ne i}^{n}( 1 - p(x_{i}))$. This is the amount by which the likelihood of each of the other users getting infected is reduced, when user $i$ invests an additional dollar.

\emph{Conditions for Investment:} If $np'(0)\prod_{j = 1, j \ne i}^{n}( 1 - p(x_{j}))R \geq -1$, it is not optimal to invest any amount in self-defense because the marginal cost of investing in defense mechanisms is relatively equal to or less than the marginal benefit of the joint reduction in risks to individuals when no investment occurs. In this case, the optimal value is a boundary investment, i.e., $x^{opt}_{i}(\overrightarrow{x}_{-i}) = 0$. If the user invests such that it has no exposure to risk, $x_{i}^{opt} = \infty$. In cases where $np'(0)\prod_{j = 1, j \ne i}^{n}( 1 - p(x_{j}))R < -1$, the optimal probabilities do not lie on the boundary and the user invests to partially eliminate risk (see Equation~(\ref{eqn:invest2})).
\subsubsection{Case 3: No Cooperation, Infection Spread} \label{sec-cc}
We assume that users do not co-operate with each other on the level of investment, i.e., users are selfish. In such a case, the optimal level of self-defense investment is the pure strategy Nash equilibria of the normal form game, $G = (N,A,u_{i}(s))$, played by the users \cite{ft}. The game consists of two players, i.e., $|N|$ = $n$; the action set of $G$ is $A = \prod_{i = 1}^{n}\times A_{i}$, where $A_{i} \,\epsilon \,[0,\infty]$, and the utility/payoff function $u_{i}(s)$ for each player $i$ is their individual final wealth, where $s\,\epsilon\,\prod_{i = 1}^{n}\times A_{i}$. The pure strategy Nash equilibria of a normal form game is the intersection of the best response functions of each user \cite{ft}.

   We define the best response function of user $i$, $x^{best}_{i}(\overrightarrow{x}_{-i})$, as
   \[x^{best}_{i}(\overrightarrow{x}_{-i})\,\epsilon\,arg max_{x_{i}}FW_{i}(x_{i},\overrightarrow{x}_{-i}),\]
   where
   \begin{equation}
   FW_{i}(x_{i},\overrightarrow{x}_{-i})= w_{0} - x_{i} - ( 1 - \prod_{j = 1}^{n}(1 - p(x_{j})))R
   \end{equation}
   Taking the first and second partial derivative of $FW_{i}(x_{i}, \overrightarrow{x}_{-i})$with respect to $x_{i}$ and equating it to zero, we obtain
   \begin{equation}
   \frac{\partial}{\partial x_{i}}(FW_{i}(x_{i}, \overrightarrow{x}_{-i})) = - 1 - p'(x_{i})\prod_{j = 1, j \ne i}^{n}(1 - p(x_{j}))R
   \end{equation}
   and
      \begin{equation}
          \frac{\partial^{2}}{\partial x_{i}^2}(FW_{i}(x_{i}, \overrightarrow{x}_{-i})) = -p''(x_{i})\prod_{j = 1,j \ne i}^{n}(1 - p(x_{j}))R < 0
      \end{equation}
   Thus, our objective function is globally concave, which implies a unique solution $x_{i}^{best}(\overrightarrow{x}_{-i})$ for each $\overrightarrow{x}_{-i}$. We also observe that a particular user $i$'s strategy complements user $j$'s strategy for all $j$, which implies that only \emph{symmetric} pure strategy Nash equilibria exist. The optimal investment for user $i$ is determined by the following equation:
   \begin{eqnarray}
   \frac{\partial}{\partial x_{i}}(FW_{i}(x_{i}, \overrightarrow{x}_{-i})) =
   - 1 - p'(x_{i})\prod_{j = 1, j \ne i}^{n}(1 - p(x_{j}))R = 0
   \label{eqn:invest3}
   \end{eqnarray}

\emph{Economic Interpretation:} The left hand side (LHS) of Equation~(\ref{eqn:invest3}) is the marginal benefit of investing in self-defense. The right hand side (RHS) of Equation~(\ref{eqn:invest3}) is the marginal cost of investing in self-defense, i.e., -1. Since the users cannot co-operate on the level of investment in self-defense mechanisms, it is not possible for them to benefit from the positive externality that their investments pose to each other.

\emph{Conditions for Investment:} If $p'(0)\prod_{j = 1, j \ne i}^{n}( 1 - p(x_{j}))R \geq -1$, it is not optimal to invest any amount in self-defense because the marginal cost of investing in defense mechanisms is greater than the marginal benefit of the joint reduction in risks to individuals when no investment occurs. In this case, the optimal value is a boundary investment, i.e., $x^{best}_{i}(\overrightarrow{x}_{-i}) = 0$. If the user invests such that it has no exposure to risk, $x_{i}^{opt} = \infty$. In cases where $p'(0)\prod_{j = 1, j \ne i}^{n}( 1 - p(x_{j}))R < - 1$, the optimal probabilities do not lie on the boundary and the user invests to partially eliminate risk (see Equation~(\ref{eqn:invest3})).

\emph{Multiplicity of Nash Equilibria:} Due to the symmetry of our pure strategy Nash equilibria and the increasing nature of the best response functions, there always exists an odd number of pure-strategy Nash equilibria, i.e., $x ^{best}_{i}(\overrightarrow{x}^{best}_{-i})$ = $x^{best}_{j}(\overrightarrow{x}^{best}_{-j})$ for all $j = 2,\ldots,n$.
\subsection{Optimal Investments Under Partial Insurance Coverage}
In this section, we analyze the situation of optimal self-defense investments when the cyber-insurance agency finds it optimal to provide partial coverage to its clients. This situation arises mainly due to conditions of information asymmetry in the insurance environment, when partial coverage is necessary to ensure a market for cyber-insurance (see Section \ref{sec-optcnr}). We only assume the realistic case of information asymmetry arising in a non-cooperative Internet environment as co-operative Internet users would want social welfare and would not generally want to hide relevant details from the cyber-insurer.

\subsubsection{Case A: No Co-operation, No Infection Spread} \label{sec-partial1}
Under partial insurance, users determine their optimal amount of self-defense investment by maximizing their expected utility of final wealth, which is \emph{not} equivalent to maximizing the expected final wealth \cite{eb}. Thus, we have to perform our analysis based on utility functions rather than based on the expected value of final wealth.

Let $U()$ be an increasing and concave utility function for each user in the network such that $U'>0$ and $U''<0$. We can determine the optimal amount of self-defense investment for each user $i$ by solving for the value of $p_{i}$ that maximizes the following constrained optimization problem:
\[arg max_{p_{i}} UFW(p_{i}) = U(w_{0} - x(p_{0} - p_{i}) - p_{i}\cdot(R - D)) \]
\[0 \leq p_{i} \leq p_{0},\]
%
%
where $UFW$ is the utility of final wealth of a user, $x(\Delta p)$, a function of the difference of $p_{0}$ and $p_{i}$, represents user $i$'s cost of reducing the risk probability from $p_{0}$ to $p_{i}$, $\Delta p = p_{0} - p_{i}$, and $0 < D < R$ is the deductible in cyber-insurance. We assume that $x$ is monotonically increasing and twice differentiable with $x(0) = 0$, $x'(0) > 0$, and $x''(0) > 0$, and $p_{i}\cdot(R - D)$ is the actuarially fair premium for user $i$'s partial insurance coverage.

\subsubsection{Case B: No Co-operation, Infection Spread} \label{sec-partial2}
Under conditions of infection spread in a non-cooperative Internet environment, user $i$'s expected utility of final wealth when a deductible of $D$ is imposed on itself is given as
\begin{equation}
UFW_{i} = UFW_{i}(p_{i},p_{-i}, D) = \alpha + \beta,
\end{equation}
where
\begin{equation}
\alpha = \prod_{i=1}^{n}(1 - p_{i})U(w_{0} - x(\Delta p_{i}) - P(D))
\end{equation}
and
\begin{equation}
\beta = 1 - \prod_{j = 1}^{n}(1 - p{j})U(w_{0} - x(p_{0} - p_{i}) - P(D) - D)
\end{equation}
We define $P(D)$ as the actuarially fair premium, and it is expressed as
\begin{equation}
P(D) = 1 - \prod_{j = 1}^{n}(1 - p{j})(R - D)
\end{equation}
Since there is spread of infection and that the Internet environment is non co-operative, we have a non co-operative game of self-defense investments between the Internet users. We denote the best response of user $i$ under a deductible as the solution to the following constrained optimization problem:
\[p_{i}^{bestD}(p_{-i}, D)\,\epsilon \, argmax_{p_{i}}UFW(p_{i}, p_{-i})\]
\[0 \leq p_{i} \leq 1,\,\forall i\]
The intersection of the best responses of the users form the set of Nash equilibria of the investment game.

\section{Comparative Study} \label{sec-cs}
In this section, we compare the optimal level of investments under full cyber-insurance coverage in the context of various cases discussed in the previous section. We emphasize here that greater the self-defense investments made by a user, better it is for the security of the whole network. Our results are applicable to Internet applications where a user has the option to be either co-operative or non-cooperative with respect to security parameters.
\subsection{Case 3 versus Case 1}
%
The following lemma gives the result of comparing Case 3 and Case 1.

\textbf{Lemma 1}. \emph{If Internet users do not co-operate on their self-defense investments (i.e., do not account for the positive externality posed by other Internet users), in any Nash equilibrium in Case 3, the users inefficiently under-invest in self-defense as compared to the case where users do not cooperate and there is no infection spread.} \\
\emph{Proof.} In Case 1, the condition for any user $i$ not investing in any self-defense is $-p'(0)R \leq 1$. The condition implies that $-1 - p'(0)\prod_{j = 1, j\ne i}^{n}(1 - p(x_{j}))R < 0$ for all $\overrightarrow{x}_{-i}$. The latter expression is the condition for non-investment in Case 3. Thus, for all users $i$, $x_{i}^{opt} = 0$ in Case 1 implies $x_{i}^{best} = 0$ in Case 3, i.e., $x_{i}^{opt}(\overrightarrow{x_{-i}^{opt}}) = x_{i}^{best}(\overrightarrow{x_{-i}^{best}}) = 0, \forall i$. The condition for optimal investment of user $i$ in Case 1 is $-1 - p'(x_{i})R = 0$. Hence, $-1 - p'(x_{i})\prod_{j = 1, j \ne i}^{n}(1 - p(x_{j}))R < 0$, for all $x_{-i}$. Thus, in situations of self-investment for user $i$, $x_{i}^{opt} > 0$ in Case 1 implies $0 \leq x_{i}^{best} < x_{i}^{opt}$, for all $x_{-i}$, in Case 3, i.e., $x_{i}^{opt}(\overrightarrow{x_{-i}^{opt}}) > x_{i}^{best}(\overrightarrow{x_{-i}^{best}}) \geq 0, \forall i$. Therefore, under non-cooperative settings, a user always under-invests in self-defense mechanisms. $\blacksquare$
\subsection{Case 3 versus Case 2}
The following lemma gives the result of comparing Case 3 and Case 2.

\textbf{Lemma 2}. \emph{Under environments of infection spread, an Internet user co-operating with other users on its self-defense investment (i.e., accounts for the positive externality posed by other Internet users), always invests at least as much as in the case when it does not co-operate.} \\
\emph{Proof.} In Case 2, the condition for any user $i$ not investing in any self-defense mechanism is $-1 - np'(0)(1 - p(0))^{n - 1}R \leq 0$. The condition also implies that $-1 - np'(0)(1 - p(0))^{n - 1}R \leq 0$. The latter expression is the condition in Case 3 for an Internet user not investing in any self-defense mechanism. Thus, for all users $i$, $x_{i}^{opt} = 0$ in Case 2 implies $x_{i}^{best} = 0$, for all Nash equilibrium in Case 3, i.e., $x_{i}^{opt}(\overrightarrow{x_{-i}^{opt}}) = x_{i}^{best}(\overrightarrow{x_{-i}^{best}}) = 0, \forall i$. The condition for optimal investment of each user $i$ in Case 2 is $-1 - np'(x_{i}^{opt}(\overrightarrow{x_{-i}^{opt}})(1 - p(x_{i}^{opt}(\overrightarrow{x_{-i}^{opt}}))^{n - 1}R = 0$. The latter expression implies $-1 - p'(x_{i}^{opt}(\overrightarrow{x_{-i}^{opt}})(1 - p(x_{i}^{opt}(\overrightarrow{x_{-i}^{opt}}))^{n - 1}R < 0$. Hence $x_{i}^{opt}(\overrightarrow{x_{-i}^{opt}}) > x_{i}^{best}(\overrightarrow{x_{-i}^{best}}) \geq 0, \forall i$. Therefore, under environments of infection spread, a user in Case 3 always under invests in self-defense mechanisms when compared to a user in Case 2. $\blacksquare$
\subsection{Case 2 versus Case 1}
The following lemma gives the result of comparing Case 2 and Case 1.

\textbf{Lemma 3.} \emph{In any $n$-agent cyber-insurance model, where $p(0) < 1 - \sqrt[n-1]{\frac{1}{n}}$, it is always better for Internet users to invest more in self-defense in a co-operative setting with infection spread than in a non-co-operative setting with no infection spread.} \\
\emph{Proof.} In Case 1, the condition for any user $i$ not investing in any self-defense is $-p'(0)R \leq 1$. The condition implies that $-1 - np'(0)(1 - p(0))^{n - 1}R \leq 0$ for all $p_{0} < 1 - \sqrt[n-1]{\frac{1}{n}}$. Thus, for all $i$, $x_{i}^{opt}(\overrightarrow{x_{-i}^{opt}}) = 0$ in Case 1 implies $x_{i}^{opt}(\overrightarrow{x_{-i}^{opt}}) \geq 0$ in Case 3 if and only if $p_{0} < 1 - \sqrt[n-1]{\frac{1}{n}}$. In situations of non-zero investment
\begin{eqnarray*}
-1 - np'(x_{i}(\overrightarrow{x}_{-i}))(1 - p(x_{i}(\overrightarrow{x}_{-i}))^{n - 1})R >
- 1 - p'(x_{i}(\overrightarrow{x}_{-i})), \forall i, \,\forall x_{i}(\overrightarrow{x}_{-i}),
\end{eqnarray*}
if and only if $p(x_{i}(\overrightarrow{x}_{-i})) < 1 - \sqrt[n-1]{\frac{1}{n}}$.  Hence,
\begin{eqnarray*}
-1 - np'(x_{i}^{opt}(\overrightarrow{x_{-i}^{opt}})(1 - p(x_{i}^{opt}(\overrightarrow{x_{-i}^{opt}}))^{n - 1})R >
- 1 - p'(x_{i}^{opt}(\overrightarrow{x_{-i}^{opt}})), \forall i,
\end{eqnarray*}
where $x_{i}^{opt}(\overrightarrow{x_{-i}^{opt}})$ is the optimal investment in Case 2. Since the expected final wealth of a user in Case 1 is concave in $x_{i}(\overrightarrow{x}_{-i})$, $x_{i}^{opt}(\overrightarrow{x_{-i}^{opt}})$ in Case 2 is greater than $x_{i}^{opt}(\overrightarrow{x_{-i}^{opt}})$ in Case 1. Thus, we infer that investments made by users in Case 2 are always greater than those made by users in Case 1 when the risk probability is less than a threshold value that decreases with increase in the number of Internet users. Hence, in the limit as the number of users tends towards infinity, the lemma holds for all $p_{0}$. $\blacksquare$

The basic intuition behind the results in the above three lemmas is that internalizing the positive effects on other Internet users leads to better and appropriate self-defense investments for users. We also emphasize that our result trends hold true in case of heterogenous network users because irrespective of the type of users, co-operating on investments always leads to users accounting for the positive externality and investing more efficiently. The only difference in case of heterogenous network user scenarios could be the value of probability thresholds i.e., $p(0)$ (this value would be different for each user in the network), under which the above lemmas hold.
%

Based on the above three lemmas, we have the following theorem.

\textbf{Theorem 1.} \emph{If Internet users cannot contract on the externalities, in any Nash equilibrium, Internet users inefficiently under-invest in self-defense, that is compared to the socially optimal level of investment in self-defense. In addition, in any Nash equilibrium, a user invests less in self-defense than if they did not face the externality. Furthermore, if $p(0) < 1 - \sqrt[n-1]{\frac{1}{n}}$, the socially optimal level of investment in self-defense is higher compared to the level if Internet users did not face the externality.}\\
\emph{Proof.} The proof follows directly from the results in Lemmas 1, 2, and 3. $\blacksquare$

The theorem implies that when negotiations could be carried out by a regulator (ex., an ISP) amongst Internet users in a cooperative setting, inevitable network externalities could be internalized and as a result users who benefit from the externality would be required to invest considerably in self-defense investments, thereby improving overall network security. The negotiations cannot not be conducted in a non-cooperative setting and as a result users would not pay for the benefits obtained from the positive externalities, thereby investing suboptimally.

\section{Optimal Cyber-Insurance Contracts} \label{sec-optcnr}
In this section, we discuss the problem of optimal insurance contracts. We make two contributions in this section: 1) we derive optimal cyber-insurance contracts under ideal insurance environments when no information asymmetry exists between the cyber-insurer and the insured and 2) we derive optimal cyber-insurance contracts under information asymmetry environments and show that a market exists for monopolistic insurance scenarios. Once optimal contracts are set by the cyber-insurance agencies, Internet users decide on their optimal self-defense investments given the optimal contracts.

\subsection{Optimal Cyber-Insurance Contracts Under No Information Asymmetry} \label{sec-oicr}
The main goal of this section is to derive optimal cyber-insurance contracts between the insurer and its clients under conditions of no information asymmetry (\emph{for perfect insurance markets}), where the insurer could have either a social welfare maximizing mindset or a profit maximizing mindset. When an insurer has a social welfare mindset, it does not care that much about making business profits as it does about insuring people so as to increase the population of users investing in self-defense mechanisms. Its hard to think of any commercial organization in the modern world who would want to provide service without thinking of profits. However, if ISPs would be a cyber-insurance agency, it would want to secure itself, being a computing and networking entity. Given that an ISP is an \emph{eyeball} and the sink for many end-user flows, it would have a strong reason to ensure high security amongst its clients as a primary objective, in order to strengthen its own security.
\subsubsection{Model} \label{sec-icm}
We assume that Internet users are uniformly distributed on the line segment [0,1], i.e., the location $p \,\epsilon\, [0,1]$ of a particular user on the unit interval denotes its probability of facing a substantial risk of size $R$. This is the risk a user faces \emph{after} some initial investments, which are precautionary efforts both in the monetary, as well as in the non-monetary sense. We assume that the ISP (or any other insurance agency) could have an estimate of this risk probability via the answers to some general questions (e.g., the type of anti-virus protection one uses, the security mindset of a user, and some basic general knowledge of Internet security) it requires its potential clients to answer before signing up for service, and from the network topology. The network topology gives information about the node degrees, which in turn helps determine the probability of each user being affected by threats.
%
Apart from the probability of facing risk, the Internet users are assumed to be homogenous in terms of their initial wealth $w$ and the size $R$ of risk faced, where a risk represents the negative wealth accumulated by a user when it is affected by Internet threats. We assume that the potential risk faced by an Internet user is less than its initial wealth $w$. Each user may buy at most one cyber-insurance policy from the insurer by agreeing to pay a premium $z$ for an insurance coverage amount $c$. The cyber-insurance company advertises only one contract to all its customers. We assume that the level of coverage is not bigger than size $R$ of risk. We also assume that the initial wealth of a user, the size of risk, the cyber-insurance premium, and the level of coverage have the same measurable units. We also account for the fact that the system does not face the information asymmetry problem.
%
%
We apply a risk-averse utility function $U_{p}(z,c)$ to Internet users, where $U_{p}(z,c)$ is defined as
\begin{equation*}
U_{p}(z,c) = \left\{
\begin{array}{rl}
w - pKR & \text{if it buys no insurance}\\
w - z - pK(R - c) & \text{if it buys insurance,}
\end{array}\right.
\end{equation*}
where $K \geq 1$ is the degree\footnote{The degree of risk aversion mentioned in this paper could be any standard risk aversion measure such as the Arrow-Pratt risk aversion measure \cite{wetzstein}.} of risk aversion of a user, assumed to be the same for all users in the network. When $K=1$, a user evaluates its loss to be exactly $R$. When $K>1$, the user adds an additional negative utility of $(K - 1)R$ for an idiosyncratic pain due to facing the risk.

We assume that the cyber-insurance agency is risk-neutral, i.e., it is only concerned with its expected profits. For an insurance policy $(z,c)$ sold to a user, the contract is worth
\begin{equation}
(1 - p)z + p(z - c) = z - pc
\end{equation}
to the insurer. Thus, the overall expected profit made by the cyber-insurance agency by providing the same insurance service to its entire geographical locality is
\begin{equation}
G(z,c) = \int_{0}^{1}(z - pc)dp
\end{equation}
Here, we use `contract' and `policy' interchangeably.
\subsubsection{Welfare Maximizing Insurance} \label{wmc}
We now determine an optimal cyber-insurance policy, $(z,c)$, a cyber-insurance agency interested in maximizing social welfare would provide to its customers. We assume here that the insurer values the welfare of each of its customers equally and is not inclined to making negative profit. We also assume that a user can decide whether to buy the policy or not, and that the insurer also has the power to decide whether to provide insurance to a customer, based on its probability of facing risk.

\noindent
{\bf Problem Formulation.}
Let the insurer offer a contract $(z,c)$. An Internet user facing a probability of risk, $p$, will want to buy cyber-insurance if $U_{p}(z,c) \geq U_{p}(0,0)$. Thus, the following condition must hold for a user to buy cyber-insurance
\begin{equation}
w - z - pK(R - c) \geq w - pKR
\end{equation}
or,
\begin{equation}
p \geq \frac{z}{Kc} = p_{L}(z,c)
\end{equation}

Therefore, a user buys insurance only if its risk probability is higher than some \emph{lower bound} $p_{L}$. The lower bound is dependent on $z, c,$ and $K$. We observe that for a fixed $K$, the lower the value of premium per unit coverage, the higher is the incentive for a user to buy cyber-insurance.

On the other hand, the cyber-insurance agency may not allow every interested user to buy insurance. There exists a particular value, $p_{H}$, of the probability of risk, for which $z = p_{H}c$. In such a case, the cyber-insurance company breaks even and the resulting $z$ is the fair premium. The insurance agency denies insurance service to users whose probability of risk is greater than $p_{H}$. Thus, $p_{H}$ is the \emph{upper bound} of the risk probability that a user requiring insurance can afford if it wants to claim insurance.

A cyber-insurer primarily interested in social welfare advertises a contract $(z,c)$ that \emph{maximizes} the total welfare of all Internet  users in its geographical locality without it making negative profits. Formally, we frame our optimization problem as follows.\\
\[argmax_{(z,c)} TW = A + B + C\]
\[subject\,to\, D,\]
where
\[A = \int_{p_{L}}^{p_{H}} [w - z - pK(R - c)]dp,\]
\[B = \int_{0}^{p_{L}} (w - pKR) dp,\]
\[C = \int_{p_{H}}^{1} (w - pKR) dp,\]
\[D = \int_{p_{L}}^{p_{H}} (z - pc)dp \,\geq 0\]
\emph{A} is the expected utility of all Internet users whose risk facing probability, $p$, lies in the interval $[p_{L}, p_{H}]$. $B$ represents the expected utility of users who have no incentive to buy insurance. The risk probability of these users lies in the interval $[0, p_{L}]$. $C$ stands for the expected utility of users who want to purchase cyber-insurance, but are denied by the insurance agency. Their risk probabilities lie in the interval $[p_{H}, 1]$. Finally, $D$ represents the constraint of the optimization problem, which states that the expected profits of the cyber-insurer are non-negative.

\noindent
{\bf Results.}
We state our results through a theorem. We note that the terms `profits' and `total user welfare' refer to the expected values of profits and social welfare. \\
\emph{\textbf{Theorem 2.}} \emph{For a welfare-maximizing cyber-insurance contract, the optimal (premium, coverage) pair is (R,R)}\emph{;} \emph{the risk probability lower bound,} $p_{L},$ \emph{equals} $\frac{1}{K}$; $p_{H} = 1$; \emph{total user welfare,} $TW,$ \emph{is} $(w - R\frac{2K - 1}{2K})$; \emph{and the insurer profit,} $P,$ \emph{equals} $R\frac{(K - 1)^2}{2K^{2}}$.\\
\emph{Proof.} We first express the risk probability bounds, $p_{L}$ and $p_{H}$, as functions of $z, K,$ and $c$. In terms of $z, K$, and $c$, $p_{L}$ is equal to $\frac{z}{Kc}$ and $p_{H}$ equals $\frac{z}{c}$. Integrating the left hand side of constraint $D$ in our optimization problem, we obtain the cyber-insurer profits as
$\frac{c}{2}\frac{z^2}{c^2}\frac{(K - 1)^2}{K^2}$. Since the profits are \emph{always} positive, the constraint $D$ is not binding on the optimization problem. Thus, our constrained optimization problem turns into the following  unconstrained one.
\[argmax_{(z,c)} P - Q + T - S,\]
where
\[P = (R - c\frac{z}{c})(\frac{K - 1}{K}),\]
\[Q = \frac{1}{2}(R - c)\frac{z^2}{c^2}(\frac{K^2 - 1}{K}),\]
\[T = w(1 - \frac{z}{c} + \frac{z}{Kc}),\]
and
\[ S = \frac{1}{2}KR(1 - \frac{z^2}{c^2} + \frac{z^2}{c^2\cdot K^2})\]
The first partial derivative of the objective function with respect to $c$ evaluates to $\frac{z^2}{c^2}$, which is a strictly non-negative quantity. Thus, the optimal value of the objective function lies at the maximum value $c$ can assume, i.e., $R$. Substituting the optimal $c$ in the objective function, we obtain a new unconstrained optimization problem of a single variable as follows.
\[argmax_{z} X - Y - Z,\]
where
\[X = (w - R\frac{z}{c})\frac{z}{c}(\frac{K - 1}{K}),\]
\[Y = \frac{w}{K}(K - \frac{z}{c}(K - 1)),\]
and
\[Z = \frac{R}{2K}(K^2 -(\frac{z^2}{c^2})(K^2 - 1))\]
The first derivative of the objective function evaluates to $R\frac{z}{c}\frac{(K - 1)^2}{K}$, which is a strictly non-negative quantity. Therefore, the optimal value of the objective function lies at $z = R$, since any premium greater than $R$ is unfair to an insurance customer and would reduce social welfare. Using substitution, the optimal (premium, coverage) pair, $(R,R)$, leads to a $p_{L}$ value of  $\frac{1}{K}$, $TW$ value of $(w - R\frac{2K - 1}{2K})$, and an insurer profit, $P$, equal to $R\frac{(K - 1)^2}{2K^{2}}$. $\blacksquare$\\
\emph{Theorem Implications:} We infer that the optimal insurance coverage in a welfare maximizing scenario is `full coverage'. For $K = 1$ the lower bound of risk facing probability, $p_{L}$, is 1, and a user buys full cyber-insurance if it is sure to face a risk, and in this case the insurer charges its client a fair premium $R$, i.e., probability of facing risk $\times$ coverage ($R$) = $R$ = premium charged. However, as the degree of risk averseness of a user increases, the value of $p_{L}$ is less than one, and a user decides to buy insurance for risks that occur with probability less than or equal to 1. Intuitively, this result makes sense as more risk averse users are more inclined to buy cyber-insurance even for risks that do not occur with probability (w.p) 1. However, for $K >1$, the insurer charges an unfair premium $R$, i.e., probability of facing risk $\times$ coverage ($R$) $< R$ = premium charged, to users who face risks that occur w.p $<$ 1, and charges a fair premium to users who are sure to face risk. Thus, the cyber-insurance agency de-incentivizes \emph{higher} risk-averse users to buy insurance when they do not face risk for sure, to prevent itself from making negative profits. The profits made by the insurance company also increase with increase in $K$, and this is true as more users buy cyber-insurance, i.e, $p_{L}$ value decreases with increase in $K$. However, the total user welfare decreases with increase in its degree of risk averseness. This is due to the fact that our utility function for each user is wealth based and a user loses more of its initial wealth with increase in its risk averseness. We emphasize here that the total user welfare is calculated by implicitly taking into account initial precautionary investments of a user. After a contract is signed between the cyber-insurer and its client, a user can decide on its optimal self-defense investments and evaluate a different utility function for welfare \cite{pg}.

\subsection{Profit Maximizing Insurance} \label{pmci}
In this section, we determine the optimal cyber-insurance policy, $(z,c)$, a cyber-insurance agency solely interested in maximizing profits (a monopolist) would provide to its customers. Similar to Section \ref{wmc}, we assume that a user can decide whether to buy the policy or not, and that the insurer also has the power to decide whether to provide insurance to a customer based on its probability of facing risk.

\noindent
{\bf Problem Formulation.}
A cyber-insurer primarily interested in making business profits chooses a contract $(z,c)$ that \emph{maximizes} its total profit over all users it services. Formally, we frame our unconstrained optimization problem as follows.
\[argmax_{(z,c)} \int_{p_{L}}^{p_{H}} (z - pc)dp,\]
subject to
\[A + B + C \ge 0,\]
where
\[A = \int_{p_{L}}^{p_{H}} [w - z - pK(R - c)]dp,\]
\[B = \int_{0}^{p_{L}} (w - pKR) dp,\]
\[C = \int_{p_{H}}^{1} (w - pKR) dp,\]
where $p_{L}$ and $p_{H}$ are defined as above.
%

\noindent
{\bf Results.}
We state our result through the following theorem. \\
\emph{\textbf{Theorem 3.}} \emph{For a profit-maximizing insurance contract, the optimal (premium, coverage) pair is} $(R\frac{K^{2}}{2K - 1}, R)$; $p_{L} = \frac{K}{2K - 1}$; $p_{H} = 1$; \emph{ and the insurer profit,} $P$\emph{,} \emph{equals} $R\frac{(K - 1)^{2}}{2(2K - 1)}$. \\
\emph{Proof.} Evaluating the integrand in the objective function, we determine the expression for overall profit as
\[P = c[\frac{z}{c}[min\{\frac{z}{c}, 1\} - \frac{z}{cK}] - \frac{1}{2}[(\{min{\frac{z}{c}, 1\}})^{2} - (\frac{z^{2}}{c^{2}K^{2}})]]\]
We observe that the expression is increasing in $c$. Thus, the cyber-insurer maximizes its profit by setting $c$ equal to $R$. When the premium per unit of coverage is less than 1, the expected overall profit is $c\frac{z^{2}}{c^{2}}\frac{(K - 1)^{2}}{2K^{2}}$, which is increasing in $\frac{z}{c}$. The increase in total profit is due to (i) increase in sales, which arises due to the increase in the range of insured individuals, i.e., the difference in the range is $p_{H} - p_{L} = \frac{z}{c} - \frac{z}{Kc} = \frac{z}{c}\frac{K - 1}{K}$, which increases with increasing premium per unit of coverage, $\frac{z}{c}$, and (ii) the mean risk probability also increasing with the premium per unit of coverage, i.e., $\int_{p_{L}}^{p_{H}}pdp = \frac{z^{2}}{c^{2}}\frac{K^{2} - 1}{2K^{2}}$, which increases with $\frac{z}{c}$. When the premium per unit of coverage is greater than 1 and $c = R$, the optimal premium is determined by equating the partial first derivative of $P$ to 0, i.e., $\frac{\partial P}{\partial (PPUC)} = \frac{z}{K^{2}}[K^{2} - \frac{z}{c}(2K - 1)] = 0$, which results in a premium $z$ equal to $R\frac{K^{2}}{2K - 1}$, where PPUC is the premium per unit of coverage. The insurer profits when PPUC is less than 1 is $R\frac{(K - 1)^{2}}{2K^{2}}$, and equals $R\frac{(K - 1)^{2}}{2(2K - 1)}$ when PPUC $\geq$ 1. Since $K^{2} > 2K -1$ for all $K\,\epsilon\,\mathbb{R}$, the cyber-insurer profits are maximized for PPUC $\geq$ 1. Substituting the values of $z$ and $c$, we get the lower bound of risk probability as $\frac{K}{2K - 1}$. $\blacksquare$\\
\emph{Theorem Implications:} We observe that full insurance coverage is the optimal insurance coverage in case of a profit maximizing scenario. Apart from the case when $K = 1$, in all other cases of $K$, the insurer charges an unfair premium to its client for a reason similar to that mentioned in the implications of Theorem 2. Taking the limit as $K$ tends to infinity, we infer that the the probability lower bound, $p_{L}$, for a user lies in the interval [0.5, 1]. The value of $p_{H}$ is obtained from the equation $R\frac{K^{2}}{2K - 1} = p_{H}R$. As for insurer profits and individual user welfare, they increase and decrease with $K$ for reasons similar to those provided in the implications of Theorem 2. We also observe that for $K = 1$, the monopolistic cyber-insurer makes zero profits. This result in in accordance with the result shown by the authors in \cite{leb3} in the context of monopolistic insurance.

\subsection{Comparison Study} \label{sec-comprstudy}
We now draw a comparison between parameters we have evaluated in both, welfare-maximizing as well as monopolistic contracts.

From the results in Sections \ref{wmc} and \ref{pmci}, we observe that the optimal premium charged by the cyber-insurers is more in the case of monopolistic insurers than in the case of social welfare-maximizing insurers, which is intuitive. For Internet users who are sure to face a risk, the monopolistic insurer charges them an unfair premium for coverage, i.e., premium $>$ coverage (except when $K$ = 1), whereas for welfare-maximizing insurers, the users who are sure to face risk are charged a fair premium as insurer is not profit maximizing. We also observe that the profits made by a monopolistic insurer are higher than its welfare-maximizing counterpart, which is also intuitive.

The total user welfare in the profit-maximizing scenario is
$w - R\frac{K^{2}(3K - 2)}{2(2K - 1)^{2}}$.
%
To compare the total user welfare in a profit-maximizing scenario with that of a welfare-maximizing scenario, we need to compare the expressions, $\frac{K^{2}(3K - 2)}{(2K - 1)^{2}}$ and $\frac{2K - 1}{K}$. Clearly, the former expression is greater or equal to the the latter for all $K \geq 1$, equality holding when $K = 1$. Therefore, the total user welfare in the case of a welfare-maximizing contract is always greater than or equal to that of a profit-maximizing cyber-insurance contract, equality holding when $K = 1$. The welfare gap for general values of $K$ is $\frac{R}{2}\cdot[\frac{K^{2}(3K - 2)}{(2K - 1)^{2}} - \frac{2K - 1}{K}]$, which is linear with $K$ - the degree of user risk averseness.

\subsection{Optimal Cyber-Insurance Contracts Under Information Asymmetry Scenarios} \label{sec-cisr}
In this section, we model realistic, i.e., imperfect, single insurer cyber-insurance markets and address two informational asymmetry problems arising between the cyber-insurer and the insured, viz., \emph{adverse selection }and \emph{moral hazard}. In adverse selection, the insurer does not know about the risk category of the user it is insuring, i.e., it does not have knowledge about whether the user is a high risk user or a low risk-user. Moral hazard results in a situation where a user behaves recklessly \emph{after} being insured, knowing the fact that it would be covered. A  cyber-insurance agency is most likely to make losses if it does not properly account for information asymmetry in its insurance contract. In this section, we design optimal cyber-insurance contracts under information asymmetry. Our analysis is suitable to scenarios of non-cooperation amongst Internet users, as we firmly believe that it is quite unlikely that users would be cooperative in regard to ensuring social welfare and at the same time behave recklessly themselves.
\subsubsection{Model} \label{sec-iamodel}
We assume two classes of users, one which has a high chance of facing risks and the other which has a low chance. We term these classes as `LC' and `HC' respectively. Let $\theta(1 - \theta)$ be the proportion of users who run a high chance(low chance) of facing risk of size $R$ respectively. However, on grounds of adverse selection the insurer cannot observe the class of any user. We consider two cases relevant to adverse selection in the Internet: 1) the insurer as well as the insured user have no knowledge about which risk class the user falls in\footnote{This situation may generally happen when the users do not provide truthful information to insurance agency questionnaires and the insurer cannot estimate the value of correlated and interdependent risks posed to users.} and 2) the insurer has no knowledge of a user's risk class but the user acquires this knowledge (through third-party agencies) after signing the contract but before it invests in self-defense investments. We assume that each user in class $i\,\epsilon\,\{LC,HC\}$ invests an amount $x_{i}$ in self-defense mechanisms after signing an insurance contract, which reduces its probability $p_{i}$ of being affected by Internet threats. We list the following mathematical properties related to our risk facing probability function $p$, for users in classes $LC$ and $HC$.
\begin{itemize}
\item $p(x)$ is a twice continuously differentiable decreasing function with $0 > p'_{LC}(x) > p'_{HC}(x)$ and $p''_{i}(x_{i}) > 0$, i.e., investments by users in class LC are more effective in reducing the loss probability than equivalent investments by users in class HC.
\item $p_{HC}(x) > p_{LC}(x)$.
\item 1 $> p_{HC}(x) \ge p_{LC}(x) > 0,\, \forall x\,\epsilon\,[0, \infty)$.
\end{itemize}
We model moral hazard by assuming that the cyber-insurer cannot observe or have knowledge about the amount of investments made by the insured. Regarding user investments, apart from the self-defense investments made by a user, we assume a certain minimum amount of base investments of value $binv$ made by an Internet user of class $i$ \emph{prior} to signing insurance contracts, without which no user can be insured. Thus $p_{i}(binv)$ is the highest chance of risk a user of class $i$ may face.

The insurance company accounts for adverse selection and moral hazard and designs an insurance contracts of the form $C = (z,c)$, for users in class $j\,\epsilon \,\{LC,HC\}$, where $z$ is the premium and $c$ is the net coverage for users. An Internet user adopts the insurance contract and invests in self-defense mechanisms to achieve maximum benefit. We measure the benefit of users of a particular risk class $i$ as a utility, which is expressed as a function of contract $C_{k}$ and self-defense investments $x_{i}$. We define the utility function for a users in risk class $i$ and facing a risk of value $R$ as an expected utility of final wealth, and it is expressed as
\begin{equation}
EU_{i}(C_{k}, x_{i}) = p_{i}(x_{i})u(w_{0} - R + c_{k}) + (1 - p_{i}(x_{i}))u(w_{0} - z_{k}),
\end{equation}
where $w_{0}$ is the initial wealth of user $i$ and $x_{i}$ is the amount of self-defense investment it makes and $u()$ is a increasing continuously differentiable function ($u'(x_{i}) > 0, u''(x_{i}) < 0$) that denotes the utility of wealth. Differentiating Equation 17 w.r.t. $x_{i}$, we get the first order condition as
\begin{equation}
-p'_{i}(x_{i})[u(w_{0} - z_{k}) - u(w_{0} - R + c_{k})] = 0
\end{equation}
The first order condition generates the optimal self-defense investment for user $i$ that \emph{maximizes} its expected utility of final wealth. In the following sections we analyze optimal cyber-insurance contracts under the presence of moral hazard when 1) neither the insurer nor the insured has any information regarding the risk class of a user and 2) the insurer does not have information regarding user class but the insured acquires information after signing the contract but before making self-defense investments.
\subsection{Neither the Insurer Nor the Insured Has Information} \label{sec-noinfo}
An Internet user does not know its risk class and therefore it maximizes its expected utility of final wealth by setting its probability of loss equal to an expected probability value of $p_{\alpha}(x) = \theta p_{HC}(x) + (1 - \theta)p_{LC}(x)$ and solving Equation 22. We assume that the values of $p_{LC}(x)$ and $p_{HC}(x)$ are common knowledge to the insurer and the insured. The cyber-insurer on the other hand, maximizes its profits by offering a contract $C_{\alpha*} = (z_{\alpha*}, c_{\alpha*})$. The optimization problem related to an insurer's profit is given as
\[argmax_{z_{\alpha}, c_{\alpha},\lambda_{\alpha},\rho_{\alpha},\rho_{0}}q_{\alpha}[1 - p_{\alpha}(x_{\alpha})z_{\alpha} - p_{\alpha}(x_{\alpha})c_{\alpha}]\]
subject to
\begin{equation}
U_{\alpha}(C_{\alpha*}, x_{\alpha*}) - U_{\alpha}(0,x_{0}) \ge 0,
\end{equation}
\begin{equation}
-p'_{\alpha}(x_{\alpha})[u(w_{0} - z_{\alpha}) - u(w_{0} - R + c_{\alpha})] = 0,
\end{equation}
\begin{equation}
-p'_{\alpha}(x_{0})[u(w_{0}) - u(w_{0} - R)] = 0,
\end{equation}
where $q_{\alpha}$ is the number of cyber-insurance contracts sold by the insurer and $x_{0}$ is the amount of self-defense investments when no insurance is purchased. $\lambda_{\alpha},\rho_{\alpha},\rho_{0}$ are the Lagrangian multipliers related to constraints 23, 24, and 25 respectively. $\alpha$ could be considered as the risk class that each user feels its in, as it does not have perfect information about whether its in class $LC$ or $HC$. Constraint 23 is the participation constraint \emph{(Individual Rationality)} stating that the expected utility of final wealth of a user is atleast as much with cyber-insurance as without cyber-insurance. Constraints 24 and 25 state that Internet users will invest in optimal self-defense investments so as to maximize their utility of final wealth, and this is in exact accordance to what the cyber-insurer wants (i.e., to avoid moral hazard). On route to solving our optimization problem, we derive the Lagrangian \cite{bv} and first order conditions, but omit it in the paper due to lack of space. Our main aim to solve the optimization problem is to only find whether the solution entails full insurance coverage or partial insurance coverage.

The optimization problem presented in this section\footnote{We also note that the optimization problems in the forthcoming sections are all examples of general principal-agent problems.} is an example of a general \emph{principal-agent} problem. The Internet users (agents) will act non-cooperatively as utility maximizers, whereas the principal's problem is to design a mechanism that maximizes its utility by accounting for adverse selection and moral hazard on the client (agent) side. Thus, the situation represents a \emph{Bayesian game of incomplete information} \cite{ft}. According to Palfrey and Srivastava \cite{ps}, there exists an \emph{incentive-compatible direct revelation mechanism} \cite{ngnp} for the problem implementable in private value models, where users do what the insurer desires (i.e., invest optimally in self-defense investments), provided the constraints in the optimization problem bind, and the users do not use \emph{weakly dominated strategies} \cite{ft} in equilibrium.

\emph{Result and Intuition:} The solution to the optimization problem in the binding case tends to \emph{full insurance} coverage as the utility function tends to become increasingly risk averse, and \emph{partial insurance} coverage otherwise. It also generates a \emph{pooling equilibrium} contract\footnote{A pooling equilibrium is one where the cyber-insurer has the same policy for both the classes (high and low risk) of users and the contract is in equilibrium.}, which is unique and entails partial cyber-insurance coverage at fair premiums. \emph{Thus, we infer that a partial insurance coverage is optimal for the cyber-insurer to provide to its clients as it accounts for the uncertainty of user risk types.} Intuitively, a pooling equilibrium works as neither the insurer nor the insured has any information on user risk type and as a result the cyber-insurer is not at a disadvantage regarding gaining risk type information relative to the Internet users. The pooling equilibrium establishes the existence of a market for cyber-insurance.

\subsubsection{Insurer Has No Information, Insured Obtains Information After Signing Contract} \label{sec-info}
In this scenario, we assume that the insurer does not have information about the risk class of a user and it cannot observe the risk class if the user obtains information from any third party agency. Since, the cyber-insurer is the first mover, it will account for the fact that users will be incentivized to take the help of a third party. We consider the case where the user may acquire information, and based on the information it decides on its self-defense investments.

Let $U_{\alpha}(C_{k},x)$ be the utility of a user in risk class $\alpha$ for a contract $C_{k}$, when it cannot observe the risk class it is in. Let $\theta U_{HC}(C_{k}, x) + (1 - \theta)U_{HC}(C_{k}, x)$ be the utility of the same user when it can get information about its risk class from a third party agency. Thus, we denote the value of gaining information to a user is $VI(C_{k})$ and its defined as
\begin{equation}
VI(C_{k}) = \theta U_{HC}(C_{k}, x) + (1 - \theta)U_{HC}(C_{k}, x) - U_{\alpha}(C_{k},x),\,\,\,0\le \theta \le 1
\end{equation}
We emphasize that $VI(C_{k})$ is zero if there is only type of risk class in the market. Now let $x_{ik}$ be the solution to Equation 18, for risk class $i$ and contract $C_{k}$. Since $p'_{LC} < p'_{\alpha} < p'_{HC}$, for contract $C_{k}$, we have $x_{LCk} > x_{\alpha k} > x_{HCk}$. Thus, $VI(C_{k}) > 0$ due to the following relationship
\begin{equation}
U_{i}(C_{k}, x_{ik}) > U_{i}(C_{k},x_{\alpha k}),\,i\,\epsilon\,\{LC,HC\}
\end{equation}
The cyber-insurer maximizes its profits by offering a contract $C_{d} = (z_{d}, c_{d})$. The optimization problem related to an insurer's profit is given as
\[argmax_{z_{d}, c_{d},\lambda_{i},\rho_{id},\rho_{i0}}\sum_{i = LC,HC}q_{i}[1 - p_{i}(x_{d})z_{d} - p_{i}(x_{d})c_{d}]\]
subject to
\begin{equation}
U_{i}(C_{d}, x_{d}) - U_{i}(0,x_{0}) \ge 0,\, i\,\epsilon\,\{ LC, HC\}
\end{equation}
\begin{equation}
-p'_{i}(x_{d})[u(w_{0} - z_{d}) - u(w_{0} - R + c_{d})] = 0,\, i\,\epsilon\,\{ LC, HC\}
\end{equation}
\begin{equation}
-p'_{\alpha}(x_{0})[u(w_{0}) - u(w_{0} - R)] = 0,\, i\,\epsilon\,\{LC, HC\}
\end{equation}
where $q_{i}$ is the number of cyber-insurance contracts sold by the insurer for class $i$ and $x_{0}$ is the amount of self-defense investments when no insurance is purchased. $\lambda_{i},\rho_{id},\rho_{i0}$ are the Lagrangian multipliers related to constraints 28, 29, and 30 respectively. Constraint 28 is the participation constraint \emph{(Individual Rationality)} stating that the expected utility of final wealth of a user is atleast as much with cyber-insurance as without cyber-insurance. Constraints 29 and 30 state that Internet users will invest in optimal self-defense investments so as to maximize their utility of final wealth (moral hazard constraints).

\emph{Result and Intuition:} The solution to the optimization problem in the binding case results in \emph{full insurance} coverage if $VI(C_{k}) = 0$ and \emph{partial insurance} coverage if $VI(C_{k}) > 0$. If $VI(C_{k}) > 0$, which is most likely the case, a user would prefer to have information on its risk class and accept contract $C_{d}$ rather than accept contract $C_{\alpha*}$ (based on utility comparisons). Our optimization problem also generates a \emph{pooling equilibrium} contract, which is unique, and entails partial coverage at fair premiums. \emph{Thus, we infer that the cyber-insurer finds its optimal to provide partial insurance coverage to its clients as it accounts for uncertainty of user risk types.} Intuitively, a pooling equilibrium works as neither the insurer nor the insured has any information on user risk type \emph{before} the user signs the contract, and as a result the cyber-insurer is not at a disadvantage with respect to gaining information on risk type relative to Internet users.

\subsubsection{Insurer Has No Information, Insured Obtains Information Prior to Signing Contract} \label{sec-info1}
In this scenario, we assume that the insurer does not have information about the risk class of a user and it cannot observe the risk class if the user obtains information from any third party agency \emph{prior} to signing the insurance contract. However, in this scenario a user that knows its risk type is at a significant advantage. Since, the cyber-insurer is the first mover, it will account for the fact that users will be incentivized to take the help of a third party. We consider the case where the user may acquire information about its risk type prior to signing the insurance contract, and based on the information it decides on the contracts and in turn its self-defense investments.
We note here that users who remain uninformed will choose contract $C_{LC}$ as its beneficial for the users to imitate the the low risk type users than be of the `expected' type.

We denote the value of gaining information to a user as $VI(C_{LC}, VI_{HC})$ and its defined as
\begin{equation}
VI(C_{LC}, C_{HC}) = \theta U_{HC}(C_{HC}, x_{HC}) + (1 - \theta)U_{LC}(C_{LC}, x_{LC}) - U_{\alpha}(C_{LC},x_{LC}),\,\,\,0\le \theta \le 1
\end{equation}

The cyber-insurer maximizes its profits by offering a contract $C_{d} = (z_{d}, c_{d})$. The optimization problem related to an insurer's profit is given as
\[argmax_{z_{i}, c_{i},\lambda_{i},\gamma_{ij},\rho_{ij},\rho_{i0}}\sum_{i = LC,HC}q_{i}[1 - p_{i}(x_{i})z_{i} - p_{i}(x_{i})c_{i}]\]
subject to
\begin{equation}
U_{i}(C_{i}, x_{i}) - U_{i}(0,x_{0}) \ge 0,\, i\,\epsilon\,\{LC, HC\}
\end{equation}
\begin{equation}
U_{i}(C_{i}, x_{i}) - U_{i}(C_{j},x_{j}) \ge 0,\, i,j\,\epsilon\,\{LC, HC\}
\end{equation}
\begin{equation}
-p'_{i}(x_{d})[u(w_{0} - z_{i}) - u(w_{0} - R + c_{i})] = 0,\, i\,\epsilon\,\{LC, HC\}
\end{equation}
\begin{equation}
-p'_{i}(x_{j})[u(w_{0} - z_{j}) - u(w_{0} - R + c_{j})] = 0,\, i,j\,\epsilon\,\{LC, HC\}
\end{equation}
\begin{equation}
-p'_{\alpha}(x_{0})[u(w_{0}) - u(w_{0} - R)] = 0,\, i\,\epsilon\,\{LC, HC\}
\end{equation}
where $q_{i}$ is the number of cyber-insurance contracts sold by the insurer for class $i$ and $x_{0}$ is the amount of self-defense investments when no insurance is purchased. $\lambda_{i},\gamma_{ij},\rho_{ij},\rho_{i0}$ are the Lagrangian multipliers related to constraints 32-36 respectively. Constraint 32 is the participation constraint stating that the expected utility of final wealth of a user is atleast as much with cyber-insurance as without cyber-insurance \emph{(Individual Rationality)}. Constraint 33 is the \emph{incentive compatibility} constraint, which states that users prefer to accept contracts that are designed to appeal to their types. Constraints 34, 35, and 36 state that Internet users will invest in optimal self-defense investments so as to maximize their utility of final wealth.

\emph{Result and Intuition:} Our optimization problem generates a \emph{separating equilibrium} contract\footnote{A separating equilibrium is one where the cyber-insurer has different insurance contracts for both the classes (high and low risk) of users and the contract is in equilibrium.}, which is unique and entails partial cyber-insurance coverage at fair premiums. \emph{Thus, even in this case, the cyber-insurer finds it optimal to provide partial insurance coverage to its clients as it accounts for the uncertainty of user risk types.} Intuitively, a separating equilibrium works as the cyber-insurer is aware of the fact that Internet users have risk type information before they lay down the contracts and thus plans different contracts for different types. In terms of optimal contracts and cyber-insurer profits, the insurer is worse off than in the no-information case because in the latter case, the insurer extracts all user surplus, whereas in the former case, it extracts full surplus from the low risk type users but only extracts partial surplus from high risk type users. The separating equilibrium establishes the existence of a market for cyber-insurance.

We have the following proposition based on the results of this section on information asymmetry cyber-insurance scenarios. \\
%
\emph{\textbf{Proposition 1:}} \emph{When neither the insurer nor the insured have any information regarding the
risk class of a user, the cyber-insurer provides full insurance coverage to
its users as their utility function becomes limiting risk averse, and
partial insurance coverage otherwise.}

\emph{If the insurer does not have any information regarding the risk class of
an insured, but the insured can gain risk class information after signing
the insurance contract, then an insured who incurs zero cost for obtaining
information finds it optimal to accept a cyber-insurance contract that
provides it full insurance coverage while it finds it optimal to accept partial
insurance coverage if the cost of obtaining information is greater than zero.}

\emph{If the insurer does not have any information regarding the risk class of
an insured, but the insured can gain risk class information before signing
the insurance contract, user welfare increases and cyber-insurer profit decreases, when compared
to the previous two cases.}

\emph{In all the three cases of information asymmetry there exists a market for cyber-insurance for single insurer
cyber-insurance environments.
}
\section{Conclusion}
In this paper, we developed a general mathematical theory of cyber-insurance contract pricing and user security investments in the Internet for single insurer cyber-insurance markets. We showed that in case of perfect insurance markets with no information asymmetry, full insurance coverage is the optimal coverage offered by the cyber-insurer, and cooperation amongst Internet users leads to better user self-defense investments w.r.t. improving overall network security. In the case of imperfect cyber-insurance environments where users are generally non-cooperative, we showed that partial insurance is the optimal cyber-insurance coverage offered by a profit-maximizing cyber-insurer. Through our models, we also show that the market for cyber-insurance exists in single cyber-insurer insurance models for both, ideal and non ideal cyber-insurance environments.

\section{Acknowledgements}
I would like to acknowledge Professor Konstantinos Psounis for his valuable comments on the area of cyber-insurance. I would like to thank the ICDCS 2010 and MAMA 2010 audience for their questions and feedback on the papers \cite{pg}\cite{mama}. In the end, I would also like to thank Professor Mung Chiang (Princeton University) and the EDGE Laboratory at Princeton University for hosting me during the summer of 2010, and giving occasional important feedback on my work on cyber-insurance.

\newpage
\bibliography{alluvion}

\begin{thebibliography}{10}

\bibitem{wik}
{\em Information Asymmetry}.
\newblock Internet Wikipedia Source.

\bibitem{ranr}
R.~Anderson.
\newblock Why information security is hard - an economic perspective.
\newblock In {\em Annual Computer Security Applications Conference}, 2001.

\bibitem{amr}
R.~Anderson and T.~Moore.
\newblock Information security economics and beyond.
\newblock In {\em Information Security Summit}, 2008.

\bibitem{rabohme}
R.~Bohme and G.~Schwartz.
\newblock Modeling cyber-insurance: Towards a unifying framework.
\newblock In {\em WEIS}, 2010.

\bibitem{ft}
D.Fudenberg and J.Tirole.
\newblock {\em Game Theory}.
\newblock MIT Press, 1991.

\bibitem{gccr}
J.~Grossklags, N.~Christin, and J.~Chuang.
\newblock Security and insurance management in networks with heterogenous
  agents.
\newblock In {\em ACM EC}, 2008.

\bibitem{hoffman}
A.~Hoffman.
\newblock Internalizing externalities of loss prevention through insurance
  monopoly.
\newblock {\em Geneva Risk and Insurance Review}, 32, 2007.

\bibitem{hv}
H.R.Varian.
\newblock {\em Microeconomic Analysis}.
\newblock Norton, 1992.

\bibitem{eb}
I.Ehrlick and G.S. Becker.
\newblock Market insurance, self-insurance, and self-protection.
\newblock {\em Journal of Political Economy}, 80(4), 1972.

\bibitem{jaw}
L.~Jiang, V.~Ananthram, and J.~Walrand.
\newblock How bad are selfish inverstments in network security.
\newblock {\em To Appear in IEEE/ACM Transactions on Networking}, 2010.

\bibitem{kmy1}
J.Kesan, R.Majuca, and W.Yurcik.
\newblock {\em The economic case for cyberinsurance: In Securing Privacy in the
  Internet Age}.
\newblock Stanford University Press, 2005.

\bibitem{kschr}
M.~Katz and C.~Shapiro.
\newblock Network externalities, competitition, and compatibility.
\newblock {\em The American Economic Review}, 75(3), 1985.

\bibitem{kmy2}
J.~Kesan, R.~Majuca, and W.~Yurcik.
\newblock Cyberinsurance as a market-based solution to the problem of
  cybersecurity: A case study.
\newblock In {\em WEIS}, 2005.

\bibitem{hh}
H.~Kunreuther and G.~Heal.
\newblock Interdependent security.
\newblock {\em Journal of Risk and Uncertainty}, 26, 2002.

\bibitem{leb3}
M.~Lelarge and J.~Bolot.
\newblock Cyber insurance as an incentive for internet security.
\newblock In {\em WEIS}, 2008.

\bibitem{leb5}
M.~Lelarge and J.~Bolot.
\newblock A local mean field analysis of security investments in networks.
\newblock In {\em ACM NetEcon}, 2008.

\bibitem{leb4}
M.~Lelarge and J.~Bolot.
\newblock Network externalities and the deployment of security features and
  protocols in the internet.
\newblock In {\em ACM SIGMETRICS}, 2008.

\bibitem{leb}
M.~Lelarge and J.~Bolot.
\newblock Economic incentives to increase security in the internet: The case
  for insurance.
\newblock In {\em IEEE INFOCOM}, 2009.

\bibitem{kmy3}
R.~P. Majuca, W.~Yurcik, and J.~P. Kesan.
\newblock The evolution of cyberinsurance.
\newblock {\em Information Systems Frontier}, 2005.

\bibitem{mybm}
R.~A. Miura-Ko, B.~Yolken, N.~Bambos, and J.~Mitchell.
\newblock Security investment games of interdependent organizations.
\newblock In {\em Allerton}, 2008.

\bibitem{ssfw}
N.Shetty, G.Schwarz, M.Feleghyazi, and J.Walrand.
\newblock Competitive cyber-insurance and internet security.
\newblock In {\em WEIS}, 2009.

\bibitem{oom}
J.~Omic, A.~Orda, and P.~V. Mieghem.
\newblock Protecting against network infections: A game theoretic perspective.
\newblock In {\em IEEE INFOCOM}, 2009.

\bibitem{pg}
R.~Pal and L.~Golubchik.
\newblock Analyzing self-defense investments in the internet under
  cyber-insurance coverage.
\newblock In {\em IEEE ICDCS}, 2010.

\bibitem{mama}
R.~Pal and L.~Golubchik.
\newblock On economic perspectives of internet security.
\newblock In {\em Workshop on Mathematical (performance) Modeling and Analysis
  (MAMA)}, 2010.

\bibitem{ps}
T.~R. Palfrey and S.~Srivastava.
\newblock Mechanism design with incomplete information.
\newblock {\em Journal of Political Economy}, 97, 1989.

\bibitem{ctm}
R.H.Coase.
\newblock The problem of social cost.
\newblock {\em Journal of Law and Economics}, 3, 1960.

\bibitem{bv}
S.Boyd and L.Vanderberghe.
\newblock {\em Convex Optimization}.
\newblock Cambridge University Press, 2005.

\bibitem{sch}
B.~Schneier.
\newblock {\em Secrets and Lies: Digital Security in a Networked World}.
\newblock John Wiley and Sons, 2001.

\bibitem{bs2}
B.~Schneier.
\newblock Insurance and the computer industry.
\newblock {\em Communications of the ACM}, 44(3), 2001.

\bibitem{bs}
B.~Schneier.
\newblock Its the economics, stupid.
\newblock In {\em WEIS}, 2002.

\bibitem{rkk}
S.Radosavac, J.Kempf, and U.C.Kozat.
\newblock Using insurance to increase internet security.
\newblock In {\em ACM NetEcon}, 2008.

\bibitem{varian}
H.~Varian.
\newblock {\em Managing Online Security Risks}.
\newblock The New York Times, June 1, 2000.

\bibitem{wetzstein}
M.~E. Wetzstein.
\newblock {\em Microeconomic Theory: Concepts and Connections}.
\newblock South Western, 2004.

\bibitem{ngnp}
Y.Narahari, D.Garg, R.Narayanam, and H.Prakash.
\newblock {\em Game Theoretic Problems in Network Economics and Mechanism
  Design Solutions}.
\newblock Springer, 2009.

\bibitem{yd}
W.~Yurcik and D.~Doss.
\newblock Cyberinsurance: A market solution to the internet security market
  failure.
\newblock In {\em WEIS}, 2002.

\end{thebibliography}
\bibliographystyle{plain}


\end{document}